\documentclass[12pt]{article}
\usepackage{amsbsy,amsmath}
\usepackage{amsfonts}
\usepackage{amssymb}
\usepackage{amscd}
\usepackage{bbm}
\usepackage{fancybox}
\usepackage{cite}
\usepackage{amsmath,amsfonts,amsbsy}
\usepackage{pstricks,pst-node}
\usepackage[small,bf,hang]{caption2}
\usepackage[linktocpage]{hyperref}
\usepackage{graphicx}
\usepackage{epsfig}
\usepackage{psfrag}
\usepackage{comment}
\usepackage{xcolor}
\usepackage[normalem]{ulem}

\psset{unit=1.3cm,linewidth=.5pt,radius=.2}  

\usepackage{multirow}                     
\usepackage{float}                          
\usepackage{lscape}                         
\usepackage{bm}

\newcommand{\bb}{\bar\beta}

\newcommand{\beq}{\begin{equation}}
\newcommand{\eeq}{\end{equation}}
\newcommand{\bi}{\begin{itemize}}
\newcommand{\ei}{\end{itemize}}
\newcommand{\bt}{\begin{tabular}}
\newcommand{\et}{\end{tabular}}
\newcommand{\bc}{\begin{center}}
\newcommand{\ec}{\end{center}}

\newcommand{\be}{\begin{equation}}
\newcommand{\ee}{\end{equation}}
\newcommand{\bea}{\begin{eqnarray}}
\newcommand{\eea}{\end{eqnarray}}
\newcommand{\ba}{\begin{array}}
\newcommand{\ea}{\end{array}}

\def\bbox{{\,\lower0.9pt\vbox{\hrule \hbox{\vrule height 0.2 cm
\hskip 0.2 cm \vrule height 0.2 cm}\hrule}\,}}
\newcommand{\dsl}{\pa \kern-0.5em /}

\font\mybb=msbm10 at 12pt
\def\bb#1{\hbox{\mybb#1}}

\makeatletter \@addtoreset{equation}{section} \makeatother

\def\slashchar#1{\setbox0=\hbox{$#1$}           
   \dimen0=\wd0                                 
   \setbox1=\hbox{/} \dimen1=\wd1               
   \ifdim\dimen0>\dimen1                        
      \rlap{\hbox to \dimen0{\hfil/\hfil}}      
      #1                                        
   \else                                        
      \rlap{\hbox to \dimen1{\hfil$#1$\hfil}}   
      /                                         
   \fi}

\begin{document}

\begin{titlepage}
\begin{center}

\vskip 1.5cm

{\Large \bf Trirefringence and the M5-brane}

\vskip 1cm

{\bf Igor Bandos${}^1$, Kurt Lechner${}^2$, Dmitri Sorokin${}^2$ and Paul K.~Townsend${}^3$} \\

\vskip 25pt

{\em $^1$  \hskip -.1truecm
\em Department of Theoretical Physics, University of the Basque Country UPV/EHU, \\
P.O. Box 644, 48080 Bilbao, Spain, \\
and IKERBASQUE, Basque Foundation for Science,
48011, Bilbao, Spain. \vskip 5pt }

{email: {\tt igor.bandos@ehu.eus}} \\

\vskip .4truecm

{\em $^2$  \hskip -.1truecm
\em I.N.F.N., Sezione di Padova \\
and  Dipartimento di Fisica e Astronomia Galileo Galilei, \\
Universit\`a degli Studi di Padova, \\
Via F. Marzolo 8, 35131 Padova, Italy  \vskip 5pt }

{email: {\tt kurt.lechner@pd.infn.it, dmitri.sorokin@pd.infn.it}}
\vskip .4truecm

{\em $^3$ \hskip -.1truecm
\em  Department of Applied Mathematics and Theoretical Physics,\\ Centre for Mathematical Sciences, University of Cambridge,\\
Wilberforce Road, Cambridge, CB3 0WA, U.K.\vskip 5pt }

{email: {\tt P.K.Townsend@damtp.cam.ac.uk}} \\

\vskip .4truecm

\end{center}

\vskip 0.5cm
\begin{center} {\bf ABSTRACT}\\[3ex]
\end{center}

The Hamiltonian formulation for nonlinear chiral 2-form electrodynamics in six-dimensional Minkowski spacetime 
is used to show that small-amplitude plane-wave perturbations of a
generic uniform constant `magnetic' background exhibit trirefringence: all three independent  
wave-polarisations have distinct dispersion relations. While two coincide
for Lorentz invariant theories, all three coincide uniquely for the chiral 2-form theory on the 
worldvolume of the M5-brane of M-theory. 
We argue that this is because, in this M-theory context, the waves propagate in a 
planar M5-M2-M2 bound-state preserving 16 supersymmetries. We also show how our results
imply analogous results for nonlinear electrodynamics in a Minkowski spacetime of  five  and four dimensions.
\bigskip

\vfill

\end{titlepage}
\tableofcontents

\section{Introduction}

The dispersion relation for an electromagnetic wave in an optically anisotropic medium is typically 
polarisation dependent; this is the phenomenon of birefringence. In theories of nonlinear electrodynamics 
(NLED), such as those arising as effective field theories for QED or SQED, a constant  uniform electromagnetic 
background can be interpreted as an optical medium for small-amplitude plane-wave disturbances, which typically
exhibit birefringence \cite{Bialynicki-Birula:1984daz}.  However, the Born-Infeld (BI) theory \cite{Born:1934gh} is an 
exception to the rule \cite{Boillat:1966eyw}; in fact, Boillat \cite{Boillat:1970gw} and Plebanski \cite{Plebanski:1970zz} 
have shown (in the closely related context of shock waves) that BI is the unique NLED with a weak-field limit for 
which there is no birefringence. There are others without a weak-field limit but, in contrast to BI,  
they are not electromagnetic-duality invariant \cite{Russo:2022qvz}. 

Implicit in the above summary of NLED birefringence results of relevance here is the choice of a four-dimensional (4D) Minkowski spacetime. In higher dimensions there are more independent polarisations. In the 5D case, for example, the gauge vector field has three independent polarisations
and one could expect to find three distinct dispersion relations for small-amplitude plane waves in a generic
constant uniform electromagnetic background; i.e. {\sl trirefringence}. As far as we are aware, this possibility has
not been investigated, presumably because there is no obvious physical motivation but it is also less interesting
from a purely theoretical perspective since one cannot expect to find any conformal limits of such theories. However, 5D NLEDs
can be obtained by dimensional reduction from nonlinear theories of 6D chiral 2-form electrodynamics, without truncation 
since the chiral restriction ensures that there are still only three independent polarisations \cite{Henneaux:1988gg}.
In this new context there are both weak-field and strong-field limits to interacting conformal chiral 2-form theories \cite{Bandos:2020hgy}. 

We have also shown in \cite{Bandos:2020hgy}, extending observations of Perry and Schwarz \cite{Perry:1996mk}, that there is a one-to-one correspondence, assuming Lorentz-invariance, between 6D chiral 2-form theories and 4D NLEDs that are electromagnetic duality invariant. This correspondence suggests that the 6D  partner to the 4D BI theory will have  special trirefringence properties. In fact, we find that it is the unique chiral 2-form electrodynamics theory for which  all three dispersion relations coincide; i.e. the unique ``zero-trirefringence'' theory.  This is an apparently stronger 
uniqueness result than could have been expected from the ``zero-birefringence'' property of the 4D BI theory because more 
conditions are needed to ensure coincidence of three dispersion relations than are needed for two. However, Lorentz invariance
is not manifest in the Hamiltonian formulation used here and, as we shall show, 6D Lorentz invariance requires, by itself, a coincidence 
of two of the three dispersion relations.  

Previous investigations into bifrefringence in the 4D NLED context have all started with a manifestly Lorentz invariant Lagrangian 
function of the electric and magnetic fields. The analogous starting point for 6D chiral 2-form electrodynamics is not immediately available because the (nonlinear) chirality condition on the 3-form field strength already implies the field equations. This difficulty can be circumvented by the inclusion of additional fields, in various ways but never without the need for some other non-manifest symmetry that imposes constraints on interactions (see \cite{Bandos:2020hgy} and references therein). As we briefly review  below,  chirality is trivially incorporated in the Hamiltonian formulation, which also applies to any theory with the same phase-space as the free-field theory, irrespective of whether it is Lorentz invariant. 

Our motivation for the investigation leading to these results comes from the importance of BI and its 6D chiral 2-form
partner in String/M-theory. The worldvolume action for the D3-brane of  IIB superstring theory is
(for suitable boundary conditions) the $\mathcal{N}=4$ supersymmetrization of the 4D BI theory \cite{Tseytlin:1996it}.
The worldvolume action for the M5-brane of M-theory \cite{Bandos:1997ui,Aganagic:1997zq} can be similarly interpreted as the $(2,0)$-supersymmetrization of the 6D chiral 2-form electrodynamics partner to the 4D BI theory \cite{Perry:1996mk}. 
In this context the 4D/6D pairing is a reflection of the String/M-theory dualities that relate the D3-brane with the M5-brane \cite{Berman:1998va,Nurmagambetov:1998gp}. We leave to the end of this article a discussion of implications
of our results in this domain. 

\section{Hamiltonian field equations}

 In the Hamiltonian formulation of 6D chiral 2-form electrodynamics, the only independent field is a 2-form
 gauge potential $A$ on the Euclidean 5-space. For time-space coordinates $\{t, x^i; i=1,\dots,5\}$, the phase-space 
 Lagrangian density takes the form \cite{Henneaux:1988gg,Bandos:2020hgy}
 \be\label{ps-Lag}
\mathcal{L} = \frac12 \dot A \cdot B - \mathcal{H}(B)\, , 
\ee
where $\dot A = \partial_t A$ and $B$ is the `magnetic' 2-form field, with components 
 \be\label{defB}
B^{ij} =(\boldsymbol{\nabla}\times A)^{ij} := \frac12 \varepsilon^{ijklm} \partial_k A_{lm} \, ,  
\ee
and $C\cdot C' = \frac12 C^{ij}C'_{ij}$ for any two 5-space 2-forms $(C,C')$. 
The Hamiltonian density $\mathcal{H}$ and the 5-vector field-momentum density 
${\bf p}$, with components
\begin{equation}\label{pi}
p_i := (B\times B)_i = \frac18 \varepsilon_{ijklm} B^{jk} B^{lm} \, , 
\end{equation}
are the Noether charge densities associated to time and space translation invariance; 
we follow here the notation and conventions of \cite{Bandos:2020hgy}:
\begin{equation}\label{gen-fe}
\dot B= \boldsymbol{\nabla} \times H\, , \qquad H := \partial\mathcal{H}/\partial B\, .  
\end{equation}
A basis for rotationally 
invariant functions of $B$ is 
\begin{equation}\label{sp2}
s= \frac12|B|^2 = \frac14 B^{ij}B_{ij} \, , \qquad 
p \equiv |{\bf p}| = |B\times B|\, , 
\end{equation}
but it is convenient to impose rotational invariance by requiring $\mathcal{H}$  to be a function 
of $(s,p^2)$, in which case
\begin{equation}
H=  \mathcal{H}_s B +2 \mathcal{H}_{p^2}\,  (\mathbf p\times B) \, ,    
\end{equation}
where, here and below, subscripts $s$ and $p^2$ denote partial derivatives  with respect to these independent variables.

Lorentz boost invariance remains non-manifest but it is a symmetry 
iff $B\times B= H\times H$ (assuming unit speed of light) \cite{Bandos:2020hgy}, and
this is equivalent to 
\begin{equation}\label{LorentzI}
\mathcal{I}:=
\mathcal{H}_s^2 + 4s \mathcal{H}_s\mathcal{H}_{p^2} + 
4p^2 \mathcal{H}_{p^2}^2 =1\, . 
\end{equation}
The choice $\mathcal{H}=s$ yields the free field theory.

\subsection{Expansion about a constant background}

For any choice of $\mathcal{H}$, the field equations are solved by $B=\bar B$, where $\bar B$ is both uniform and constant.
We may expand the field equation for $B$ about this `background' solution, which can then be viewed as a stationary 
homogeneous `optical'  fluid medium of energy density $\bar{\mathcal{H}} = \mathcal{H}(\bar B)$ and momentum 
density $\bar {\bf p}$. We may consider perturbations about this background by setting $A= \bar A + {\rm a}$, 
where  $\boldsymbol{\nabla}\times \bar A = \bar B$ (note that $\bar A$ cannot be uniform for non-zero $\bar B$). This implies that
\begin{equation}
B= \bar B + {\rm b}\, , 
\qquad {\rm b}= \boldsymbol{\nabla} \times {\rm a}\, , 
\end{equation}
and hence that $\partial_i b^{ij} \equiv 0$. Expanding the field equation \eqref{gen-fe} to first order in $b$ we find
that
\begin{equation}\label{linearb}
\dot b = \boldsymbol{\nabla} \times h(b) \, , 
\end{equation}
where $h(b)$ is a two-form depending linearly on $b$: 
\begin{eqnarray}\label{Udef}
h_{ij} &=& Q\, b_{ij}+2\bar{\mathcal{H}}_{p^2}\left(\bar B b\bar B + \bar B^2 b + b\bar B^2\right)_{ij}\nonumber\\
 &&+ \left[(Y + 4\bar sX)(\bar B\cdot b) + 2X (\bar B^3\cdot b)\right] \bar B_{ij} \nonumber\\
&& +  2\left[X (\bar B\cdot b)  + 2\bar{\mathcal{H}}_{p^2p^2} (\bar B^3\cdot b)\right] (\bar B^3)_{ij}
\end{eqnarray}
for coefficient functions
\begin{equation}
\begin{aligned}
Q =& \bar{\mathcal{H}}_{s}+ 4\bar s \bar{\mathcal{H}}_{p^2}\, , \\
X =& \bar{\mathcal{H}}_{sp^2} + 4\bar s\, \bar{\mathcal{H}}_{p^2p^2} \, , \\
Y=&4 \bar{\mathcal{H}}_{p^2} + \bar{\mathcal{H}}_{ss} + 4\bar s\, \bar{\mathcal{H}}_{s p^2}\, . 
\end{aligned}
\end{equation}
In what follows we shall omit the bars on the background values of $(s,p)$, 
and on $\mathcal{H}$ and its derivatives; it should be clear from the context when we are considering a constant uniform background and when we are considering generic field configurations. However, we will retain the $\bar B$ notation for the background value of $B$. 

\subsection{Plane waves on the background} 

For plane-wave solutions of \eqref{linearb} with angular frequency $\omega$ and wave 5-vector ${\bf k}$,
the amplitudes $\text{b}_{ij}$ satisfy 
\begin{equation}\label{2eqs}
-\omega \text{b} = {\bf k} \times h(\text{b})\, , 
\qquad  k_i \text{b}^{ij}=0\, , 
\end{equation}
but the second of these equations is implied by the first unless $\omega=0$. The first equation can 
be written as a $10\times10$ matrix equation of the form $M(\omega,k)\underline{\text{b}}=0$, 
where the components of $\underline{\text{b}}$ are the ten independent components of the 2-form $\text{b}$. 
Each non-zero solution corresponds to a zero of $\det M(\omega,k)$, a 10th-order polynomial in $\omega$.

 Using the $O(5)$ rotation/reflection symmetry, we may choose the 5-space axes such that the only non-zero components of $\bar B$ are 
\begin{equation}
\bar B^{12} = - \bar B^{21} = \text{B}_1\, , \qquad 
\bar B^{34} = -\bar B^{43} = \text{B}_2\, , 
\end{equation}
for constants $\text{B}_1\ge \text{B}_2\ge0$, so that $\text{B}_1\text{B}_2 = p$. This canonical form for $\bar B$ preserves an $SO(2) \times SO(2)$ subgroup of $O(5)$, which we may use to set 
\begin{equation}
k_2=k_4=0\, . 
\end{equation}
The matrix $M(\omega,k)$ is now block diagonal if we choose the first four components of $\underline{\text{b}}$ to 
be $(\text{b}_{24},\text{b}_{13}, \text{b}_{15}, \text{b}_{35})$, so its determinant must factorise: $\det M = \Delta_4\Delta_6$. 
One finds that
\begin{equation}
\Delta_4 = \omega^2 P_2(\omega)\, , 
\qquad \Delta_6= \omega^2 P_4(\omega)\, , 
\end{equation}
for polynomials $P_2$ and $P_4$ of, respectively, second and fourth order in $\omega$. The four linearly independent solutions with $\omega=0$ are eliminated by the four conditions $k_i \text{b}^{ij}=0$, so 
plane-wave solutions correspond to zeros of either $P_2$ or $P_4$.

A calculation yields 
\begin{equation}\label{P2g}
P_2 = (\omega + 2k_5 p\,\mathcal{H}_{p^2})^2 - \chi \,   
\ee
where
\be
\chi= \mathcal{H}_s^2 k_5^2  + \mathcal{H}_s\left(Q_1 k_1^2 + Q_2 k_3^2\right)\,  , 
\ee
with 
\be
Q_\alpha = \mathcal{H}_s + 2\text{B}_\alpha^2 \mathcal{H}_{p^2}  \qquad (\alpha=1,2).
\end{equation}
The dispersion relation for one polarisation is therefore $P_2=0$, 
with $P_2$ given by \eqref{P2g}. As expected, it reduces to $\omega^2 = |{\bf k}|^2$ in the free-field limit. More generally, it depends on the direction of the wave-vector because of the term in \eqref{P2g} with the factor of $k_5p$ (i.e.  ${\bf k} \cdot{\bf p}$).  

The remaining two dispersion relations must be obtained from the condition $P_4=0$. A calculation yields
\be\label{P4new}
P_4 = \left\{ \left[\omega + 
k_5 p\left(2\mathcal{H}_{p^2} + \Lambda\right)\right]^2  
- \chi'\right\}P_2  + \Upsilon  k_1^2k_3^2\, ,
\ee
where 
\be
\Lambda = 2\mathcal{H}_{p^2} + \mathcal{H}_{ss} +
4s \mathcal{H}_{sp^2} + 4p^2\mathcal{H}_{p^2p^2}  
\ee
and
\be
\chi'= \Xi_1\Xi_2 k_5^2 + \Xi_1 Q_2 k_1^2 + \Xi_2 Q_1 k_3^2 \, , 
\ee
for the additional coefficient functions
\be
\begin{aligned}
\Xi_\alpha =& \mathcal{H}_s + 2s \mathcal{H}_{ss}  + 4p^2 
\mathcal{H}_{sp^2}  \\
&+ \text{B}_\alpha^2 (\Lambda - 4s \mathcal{H}_{sp^2} 
- 2\mathcal{H}_{ss})\, , 
\end{aligned}
\ee
and finally
\be\label{Ups}
\Upsilon = N_1 N_2 - Q_1Q_2\, p^2\Lambda^2  
\ee
with
\be
\begin{aligned}
N_1 &= Q_2\Xi_1 -  \mathcal{H}_s Q_1\, ,\\
N_2 &= Q_1\Xi_2 -  \mathcal{H}_s Q_2\, .
\end{aligned}
\ee

\subsection{Zero trirefringence conditions}

The conditions required for all three dispersion relations to 
coincide is $P_4=P_2^2$. From  \eqref{P4new}
we see that $P_2$ is a factor of $P_4$ for generic ${\bf k}$ 
only if $\Upsilon=0$. The other factor is also
$P_2$ only if both $\Lambda=0$ and $\chi'=\chi$ for all ${\bf k}$, 
which requires only that $N_1=N_2=0$
since these two relations imply $\Xi_1\Xi_2 = {\mathcal{H}}^2_s$. 
Moreover, the three relations 
\be\label{no-tri1}
N_1=N_2= \Lambda =0\, 
\ee
imply $\Upsilon=0$, so these three relations are the necessary and sufficient conditions for  
coincidence of all three dispersion relations. They may be simplified by the observation that 
\bea\label{Np1}
& \!\!\!\! \!\!\!\! N_1+N_2  = 
2\left(s\mathcal{H}_s+2 p^2\mathcal{H}_{p^2}\right)\Lambda 
-8\left(s^2-p^2\right)\Lambda_1, &\\
 &\!\!\!\!\! \!\!\!\! \!\!\!\!  
 N_1-N_2 = 2 \sqrt{s^2-p^2} \left(4s\Lambda_1 + 8p^2\Lambda_2 - \mathcal{H}_{s} \Lambda\right),& \label{Nm1}
\eea
where 
\begin{equation}
\begin{aligned}\label{first2}
 \Lambda_1:=&
 \mathcal{H}_s \mathcal{H}_{sp^2} - 
 \mathcal{H}_{p^2}\mathcal{H}_{ss}  \\
 \Lambda_2:=&
 \mathcal{H}_s\mathcal{H}_{p^2p^2} - 
 \mathcal{H}_{p^2}\mathcal{H}_{sp^2} \, .   
\end{aligned}
\end{equation}
This shows that equations \eqref{no-tri1} are jointly equivalent to the following 
three ``zero trirefringence'' conditions:
\be\label{zero-tri2}
\Lambda_1=0\, , \qquad  \Lambda_2 =0\, , \qquad  \Lambda=0\, . 
\ee
The first two of these equations are trivially solved if $\mathcal{H}_{p^2}=0$, but then the third requires $\mathcal{H}$ to be a linear function of $s$. Excluding this free field case, we may assume that $\mathcal{H}_{p^2}\ne0$ and then define a new function $T(s,p^2)$ by the relation 
\be\label{Wronskian}
\mathcal{H}_{s}  = 2T \mathcal{H}_{p^2}\, . 
\ee
Using this in the equations $\Lambda_1=\Lambda_2=0$ we find that
\be\label{THp2=0}
T_s \mathcal{H}_{p^2}^2 = T_{p^2} \mathcal{H}_{p^2}^2 =0\,,   
\ee
from which we conclude that $T$ is a constant. Using this fact to simplify the $\Lambda=0$ condition, we then  find the following simple second-order ODE for ${\mathcal{H}}$ 
as a function of $p^2$:
\be\label{M5Ham}
\mathcal{H}_{p^2} + 2\left(T^2 +2Ts +p^2\right) 
\mathcal{H}_{p^2p^2}=0\, .  
\ee
The general solution has two integration constants. One is fixed by requiring positive energy and a choice of energy scale. 
The other is then fixed by requiring zero vacuum energy. The result is  
\be\label{HamM5}
\mathcal{H} = \mathcal{H}_{\text{M5}} := 
\sqrt{T^2 + 2Ts + p^2} -T \, . 
\ee
This is the Hamiltonian density for the chiral 2-form theory on the M5-brane \cite{Bergshoeff:1998vx,Townsend:2019ils}; for brevity we shall call it the
`M5' theory. It is actually a family of theories labelled by the constant $T$ (the M5-brane tension) which has dimensions of energy density, and the free-field theory is included as the $T\to\infty$ limit. As already mentioned this `M5' theory is the 6D partner to the 4D BI theory. It would be of interest to see whether there is a generalisation to 6D chiral 2-form dynamics of the recent characterisation of zero-birefrigence NLEDs as those with a Lagrangian satisfying a particular ``$T\bar T$-like flow equation'' \cite{Ferko:2023ruw}.

To summarise: within the class of chiral 2-form electrodynamics invariant under rotations and time-space translations, and with the same phase space
as the standard free-field theory, only the one-parameter `M5' family exhibits ``zero-trirefringence''. For this 
exceptional family, the one dispersion relation  for the three independent wave-polarizations is  found from 
setting $\mathcal{H} = \mathcal{H}_{\text{M5}}$ in \eqref{P2g} and then setting the resulting
expression for $P_2$ to zero; this yields
\be\label{M5disp}
\left[\omega +\frac {{\bf k}\cdot{\bf p}}{T_{\rm eff}}\right]^2  = \frac{T^2 |{\bf k}|^2- T|\bar B{\bf k}|^2 }{T_{\rm eff}^2} \, , 
\ee
where
\be\label{Teff}
T_{\rm eff}= \sqrt{T^2+2T\bar s + \bar p^2} \, . 
\ee
Here we revert to the bar notation for background fields as a reminder that $T_{\rm eff}$ (which will play a role later) is constant.  
In the $T\to \infty$ (weak-field) limit this dispersion relation reduces to $\omega^2=|{\bf k}|^2$, as expected. 

We may also take the $T\to0$ (strong-field) limit for which $\mathcal{H}=p$. This defines an interacting 
{\sl conformal} 6D chiral 2-form electrodynamics theory \cite{Gibbons:2000ck,Townsend:2019ils}; its 4D partner is 
Bialynicki-Birula electrodynamics \cite{Bialynicki-Birula:1984daz,Bialynicki-Birula:1992rcm}. All constant uniform background 
solutions now have $p\not = 0$, and \eqref{M5disp} reduces to the linear dispersion relation 
$\omega +{\bf k}\cdot {\bf n}=0$, where ${\bf n} = {\bf p}/p$. In this case $b$ is a Fourier component of the first term
in an expansion about $B=\bar B$ of an exact solution of the full field equations of the form $B= B_\perp(t- {\bf x}\cdot{\bf n}, {\bf x}_\perp)$, 
where $n_iB^{ij}_\perp=0$ and ${\bf n} \cdot {\bf x}_\perp=0$ for fixed direction ${\bf n}$. 

\subsection{Lorentz invariance}

Surprisingly, the above results were obtained without the use
of the Lorentz invariance condition \eqref{LorentzI}, which is actually
a consequence of the zero-trirefringence conditions \eqref{zero-tri2}.
We shall now show that the Lorentz invariance condition by itself
restricts trirefringence to birefringence; i.e. it implies that two of the three independent dispersion relations coincide. We begin with the observation that
\bea\label{Isp2}
\mathcal{I}_s & = & \ 2 (\mathcal{H}_s\Lambda-2s\Lambda_1-4p^2\Lambda_2)\,,\nonumber\\
\mathcal{I}_{p^2} & =& \  2(\mathcal{H}_{p^2}\Lambda+\Lambda_1+2s\Lambda_2)\, . 
\eea
Next, we observe that \eqref{Nm1} may be rewritten as 
\be\label{Npm2}
N_1-N_2 = 2 \sqrt{s^2-p^2} \left(\mathcal{H}_{s} \Lambda - \mathcal{I}_s\right)\, . 
\ee
Now, using \eqref{Npm2} together with 
{\eqref{Np1} and \eqref{Isp2} we obtain
\bea
N_1N_2 &\equiv&  \frac 14 [(N_1+N_2)^2-(N_1+N_2)^2]\\
&=&8(s^2-p^2)p^2\left(\Lambda_2\mathcal I_s-\Lambda_1\mathcal I_{p^2}\right)+\mathcal I p^2\Lambda^2\,.\nonumber 
\eea
Substituting this expression into \eqref{Ups}, and using the identity
\be
Q_1Q_2 \equiv \mathcal{I}\, , 
\ee
where $\mathcal{I}$ is the expression defined in \eqref{LorentzI}, 
we deduce that 
\be\label{Ups'}
\Upsilon = 8(s^2-p^2)p^2\left(\Lambda_2\mathcal I_s-\Lambda_1\mathcal I_{p^2}\right) \, . 
\ee
This result shows that $\Upsilon=0$ for any $\mathcal{H}$ such that $\mathcal{I}=1$,  i.e. any Lorentz invariant theory. It then follows from \eqref{P4new} that Lorentz invariance implies $P_4=P'_2P_2$, where $P'_2(\omega)$ is another quadratic polynomial in $\omega$. Thus, two of the three independent polarisations have coincident
dispersion relations for generic ($P_2'\ne P_2$) Lorentz invariant theories 
while $P_2'=P_2$ uniquely for the `M5' case.

\section{Relation to M-theory}

It is natural to wonder whether there is some M-theory explanation for the zero-trirefringence property 
of the `M5' chiral 2-form theory. In the context of the M5-brane worldvolume dynamics, the Minkowski vacuum for the `M5' theory 
is a planar static M5-brane, and perturbations about it are propagated by the free-field equations of a (2,0)-supersymmetric 6D field theory; its on-shell supermultiplet includes the three polarisation modes of the `M5' chiral 2-form electrodynamics and five
others, one for each of the five scalars representing transverse fluctuations of the planar M5-brane in an 11-dimensional 
space-time \cite{Gibbons:1993sv}.  It might appear that some of 
the 16 supersymmetries of this (2,0)-supermultiplet must be broken when constant uniform background fields are introduced on the M5-brane, but this is not necessarily the case, as we now explain.  

It was shown in \cite{Sorokin:1997ps} that a static planar M5-brane with constant uniform 3-form field strength is $\frac12$-supersymmetric; i.e. it preserves 16 of the 32 supersymmetries of the M-theory 11D Minkowski vacuum, {\sl independently of the strength of the `background' 3-form field}. 
If the skew eigenvalues $\{\text{B}_\alpha; \alpha=1,2\}$ of the background `magnetic' 2-form are identified as ``dissolved'' M2-branes with charges 
\be\label{M2charges}
\zeta_\alpha = \sqrt{T}\,  \text{B}_\alpha
\ee
then the $\frac12$ supersymmetry is also implied by the supertranslation algebra associated to the M5-brane worldvolume dynamics
provided that \cite{Sorokin:1997ps,Bergshoeff:2000qn}
\be\label{P5}
P_5 = \zeta_1\zeta_2/T\, , 
\ee
which is the background field-momentum.  The `effective' M5-brane tension (i.e. total energy density) of these bound states is
\be\label{effT}
P^0 = \sqrt{T^2 + \zeta^2_1 +\zeta^2_2 + P_5^2} \, . 
\ee
The construction in \cite{Bergshoeff:2000qn} of 11D supergravity solutions sourced by these M5-M2-M2 ``bound states'' (generalizing the simpler M5-M2 $\frac12$-supersymmetric solution of \cite{Izquierdo:1995ms}) confirmed their 
$\frac12$ supersymmetry. They are related by String/M dualities to the D2-D0-F1 ``supertube'' bound states of IIA superstring theory  \cite{Mateos:2001qs,Emparan:2001ux} but with a {\sl planar} D2-brane for which the generic $\tfrac14$ supersymmetry is
enhanced to $\tfrac12$ supersymmetry \cite{Mateos:2001pi}. 

Using \eqref{M2charges} and \eqref{P5}, and reverting to the bar notation for background fields, we may rewrite \eqref{effT} as 
\be
P^0 = \sqrt{T^2 + 2T\bar s + \bar p^2} = T_{\rm eff}\, , 
\ee
which is the expression of \eqref{Teff}, now interpreted as the effective M5-brane tension for the
M5-brane plus $\bar B$ background, which we can view as a new worldvolume `vacuum' preserving all 16 supersymmetries. We should then re-normalize the M5-brane vacuum energy to be zero when $B=\bar B$. This means that we should replace 
$\mathcal{H}_{M5}$ by
\be
\mathcal{H}'_{M5} = \sqrt{T^2 + 2Ts +p^2} - T_{\rm eff}\, . 
\ee
Now, in the expansion about the $B=\bar B$ background, the energy is zero when $b=0$. We thus expect the 
field equations \eqref{linearb} to be part of a larger set of 
equations for disturbances of a planar M5-M2-M2 bound state configuration preserving 16 supersymmetries. This leads us to conjecture that the 
zero-trirefringence property of the `M5' theory is a consequence of its
unique status as a consistent truncation of the maximally-supersymmetric 6D field theory found from expansion of the full M5-brane 
dynamics about a novel 1/2-supersymmetric vacuum.  

\section{Implications for 5D and 4D NLED} 

We conclude with a brief explanation of how our results for 6D chiral 2-form electrodynamics 
imply analogous results for 5D and 4D NLEDs by means of dimensional reduction.  As we used 
symmetries preserved  by the 6D $B=\bar B$ background solution to choose the wave-vector 
of perturbations to have zero $k_2$ and $k_4$ components, we will first take all fields to be
independent of $x^2$, to get 5D results, and then of both $x^2$ and $x^4$, to get 4D results. In the former case the 5-space 2-form 
$A= \frac12 dx^i\wedge dx^j A_{ij}$ can be written as
\be
\frac12 dx^a\wedge dx^b \bb{A}_{ab} + dx^2\wedge dx^a V_a\, , \quad
(a,b = 1,3,4,5). 
\ee
Correspondingly,  
\be
\begin{aligned}
B^{ab} &= \frac 12 \varepsilon^{abcd} F_{cd}\,, \quad  (F_{ab} =2 \partial_{[a} V_{b]}) \\
B^{a2} &= \frac12\varepsilon^{abcd} 
\partial_b \bb{A}_{cd} =: D^a\, .
\end{aligned}
\ee
The gauge-invariant 4-space fields are therefore the two-form field strength $F_{ab}$} and a divergence-free 4-vector field $D^a$ that can 
be `promoted' to an unconstrained 4-vector field by introducing a Lagrange multiplier field $V_0$ to impose the constraint $\partial_aD^a=0$.  One then finds (ignoring total derivative terms) that 
\be
\frac 12 \dot A \cdot B = E_a\, D^a\,  \qquad 
(E_a = \partial_a V_0 - \dot V_a) \, .
\ee
This is the `symplectic' term in the phase-space Lagrangian density \eqref{ps-Lag} reduced to a 5D NLED. Its Hamiltonian is a function of $(s,p^2)$ but now
\be
s = \frac12\left[|D|^2 + |F|^2\right]\, .  
\ee
and, since the components of ${\bf p}$ in \eqref{pi} are now
\be
p_2 =\frac18 \varepsilon^{abcd} F_{ab}F_{cd}\, , \qquad p_a = F_{ab} D^b\, , 
\ee
we also have
\be
p^2   = \det F + |FD|^2\, .
\ee
For the special class of 5D NLEDs with Hamiltonian densities that are functions only of $(s,p^2)$, 
our  6D results imply that the unique zero-trirefringence family has a Hamiltonian density that is formally the same as
$\mathcal{H}_{M5}$ of \eqref{HamM5}, but this is now equivalent to 
\be\label{5DBI}
\sqrt{ T^2 \det\left(\bb{I}_4 + F/\sqrt{T}\right) + T|D|^2 + |FD|^2} -T\, . 
\ee
This is the 5D BI Hamiltonian density, as can be seen from previous results on the 
Hamiltonian dynamics of the bosonic worldvolume fields for Dp-branes \cite{Lindstrom:1997uj,Gauntlett:1997ss}. 

We may now further dimensionally reduce by taking all fields to be independent of $x^4$. In this case $V_4$ becomes a scalar field, 
with canonical conjugate $D^4$, and if we truncate by setting to zero this conjugate pair then we are left with the 3-vector ${\bf D}$ (conjugate
to ${\bf V}$) and the 3-space restriction of $F$, which is the Hodge dual of the magnetic 3-vector field ${\bf B}$. The phase space Lagrangian 
density is now that of a 4D NLED with Hamiltonian density $\mathcal{H}$ that is again a function of $(s,p^2)$, but now 
\be
s = \tfrac12\! \left(|{\bf D}|^2 + |{\bf B}|^2\right)\, , \qquad p^2 = |{\bf D}\times{\bf B}|^2\, ,
\ee
which implies that $\mathcal{H}$ is electromagnetic duality invariant. Our 6D trirefringence results now imply that all zero-birefringence
4D NLEDs in this class are members of the family for which $\mathcal{H}(s,p^2)$ is formally the same as $\mathcal{H}_{M5}$ of \eqref{HamM5}, 
but this now defines the 4D BI theory. In other words, the 4D BI theory is the unique duality invariant zero-birefringence NLED, in agreement with \cite{Russo:2022qvz} and with an earlier conclusion of \cite{Deser:1998wv} based on (what appear to us to be) slightly different premises; the main novelty here is that we find it in Hamiltonian form, and by dimensional reduction of the `M5' theory of 6D chiral 2-form electrodynamics. 

From the 6D perspective, the truncation described above (following dimensional reduction) amounts to setting to zero the components 
$(B_{24}, B_{13}, B_{15}, B_{35})$ of $B$,  only two of which are independent because  of the identity $\partial_i B^{ij}\equiv 0$. This truncation also removes the two linearly independent combinations of the $(\text{b}_{24},\text{b}_{13}, \text{b}_{15}, \text{b}_{35})$ perturbations of $B$ about the $\bar B$ background. 
In 6D these were the amplitudes describing the phase space for the single polarisation mode with dispersion relation $P_2=0$ (as required for
consistency of the truncation). In 4D they are the amplitudes for the `extra' scalar field in the 4D $\mathcal{N}=4$ Maxwell supermultiplet relative to the 6D (2,0) antisymmetric tensor supermultiplet. 

In the String/M-theory context, the `dissolved' pair of orthogonal M2-branes (representing the $\bar B$ background  on a static planar M5-brane) becomes a `dissolved' orthogonal F1-D1 pair of IIB strings, i.e. a constant uniform background of orthogonal $({\bf D},{\bf B})$ fields, which generate the momentum ${\bf D}\times {\bf B}$ needed to preserve all 16 supersymmetries of a static planar D3-brane in the absence of the electromagnetic background fields.

\section*{Acknowledgements}
IB and DS have been partially supported  
by Spanish AEI MCIN and FEDER (ERDF EU) under grant PID2021-125700NB-C21 and by the Basque Government Grant IT1628-22. 
PKT has been partially supported by STFC consolidated grant ST/T000694/1.

\providecommand{\href}[2]{#2}\begingroup\raggedright

\end{document}